\documentclass[pra,twocolumn,preprintnumbers,amsmath,amssymb,superscriptaddress,showpacs,longbibliography]{revtex4-2}

\RequirePackage{fix-cm}
\RequirePackage{amsmath}

\usepackage[T1]{fontenc} 
\usepackage[utf8]{inputenc} 

\usepackage{graphicx}
\usepackage{latexsym}
\usepackage{amsmath}
\usepackage{graphics}
\usepackage{amssymb}
\usepackage{layout}
\usepackage{verbatim}
\usepackage{amsfonts,epsfig}
\usepackage{soul}
\usepackage[dvipsnames,table]{xcolor} 
\usepackage{hyperref}
\usepackage{tikz}
\usepackage[compat=1.1.0]{tikz-feynman}
\usepackage{color}
\hypersetup{colorlinks, linkcolor={blue!90!black}, citecolor={blue!90!black}, urlcolor={blue!90!black} }
\usetikzlibrary{quantikz}

\newcommand{\beq}{\begin{equation}}
\newcommand{\eeq}{\end{equation}}
\newcommand{\bea}{\begin{eqnarray}}
\newcommand{\eea}{\end{eqnarray}}

\newcommand{\tr}{\hbox{Tr}}

\usepackage{color}

\usepackage{ulem} 

\begin{document}

\title{Entanglement with a mode observable via a tunable interaction with a qubit
}

\author{Ma{\l}gorzata Strza{\l}ka}
\affiliation{FZU - Institute of Physics of the Czech Academy of Sciences, Na Slovance 2, 182 00 Prague, Czech Republic}
\affiliation{Faculty  of  Mathematics  and  Physics, Charles University, Ke Karlovu 5,
121 16 Prague, Czech Republic}
\author{Radim Filip}
\affiliation{
Department of Optics, Palack{\'y} University, 17. Listopadu 12, 771 46 Olomouc, Czech Republic}
\author{Katarzyna Roszak}
\affiliation{FZU - Institute of Physics of the Czech Academy of Sciences, Na Slovance 2,  182 00 Prague, Czech Republic}

\date{\today}

\begin{abstract}
We study the possibility of detection of ``spin-boson'' entanglement by qubit only measurements. Such entanglement is impossible to detect by previously proposed schemes that involve a fixed system-environment
interaction, because of 
inherent symmetries within the coupling and the initial state of the environment. We take advantage
of the possibility of tuning of qubit-environment coupling, that is available in some qubit realizations.
As an example we study a superconducting transmon qubit interacting
with a microwave cavity, which is one of such systems and is, 
furthermore,
essential in the context of quantum information processing. We propose suitable Hamiltonian parameters for the preparation 
and measurement phases of the detection scheme that allow for an experimental test, and verify that the reported signal is 
nonnegligibly large still at finite temperatures.	
\end{abstract}
\maketitle

\section{Introduction }
A coupling of an environment to quantum systems has a dichotomic role, simultaneously positive and negative.
Entanglement created between a qubit (or qubits) and its 
environment is responsible for decoherence \cite{zurek03}, but
can also be a resource for efficient operation of quantum algorithms. 
Specifically, the quantum nature of 
an environment
can be used for disspative state preparation, measurement and stabilization \cite{harrington22}, quantum error correction \cite{gregoratti03}, and it also
has the power to positively
influence the operation of algorithms \cite{roszak23}
which is inaccessible for classical noise.
This is because quantum correlations
enhance the effect that operations and measurements performed on qubit(s) can have on the environment
due to information transfer between the two subsystems \cite{roszak17}. 
This in turn may lead to classically unpredictable effects and qubit decoherence curves that cannot be modeled in separation from the actual environmental state
\cite{roszak21,harlender22,roszak23}.

Without energy relaxation, the qubit can undergo only pure dephasing by the coupling to environmental modes
(environmentally induced transitions between qubit pointer states do not occur).
Such decoherence can be the result either of entanglement generation or of the build-up of classical 
correlations \cite{eisert02,pernice11}. 
Qubit-environment entanglement (QEE) which accompanies a pure-dephasing (PD) interaction can be detected experimentally \cite{bizzarri25}, because for such couplings 
the state of the environment is
qualitatively different when entanglement with the qubit is generated \cite{roszak15,roszak18}. 
Although the direct measurement of the environment is challenging, it is possible to transfer the
information about environmental states (and thus entanglement) back to the qubit via subsequent qubit operations 
and measurements. Therefore QEE can be witnessed by probing the qubit
without direct access to the environment \cite{roszak19a,rzepkowski21,strzalka21}.

The detection schemes of Refs \cite{roszak19a,rzepkowski21,strzalka21} assume that there is no control over the qubit-environment interaction and the interaction is completely fixed.
In this case, the necessary condition for the QEE detection to work
requires that operators that describe the evolution of the environment conditional on the pointer state of the 
qubit do not commute. Otherwise the information cannot be transferred from the environment into the qubit,
even though entanglement does leave a measurable trace in the environment itself.
Additionally, for QEE to be detected, the information about entanglement cannot be lost when the environment is traced out. This crucially depends on the symmetries present in the qubit-environment (QE) state.
A canonical example of a mixed QE system for which decoherence is accompanied by entanglement generation,
but the entanglement cannot be detected via the schemes of Refs \cite{roszak19a,rzepkowski21,strzalka21},
is the generic spin-boson model. A detailed discussion of the limitations of such schemes is found in the Appendix of Ref.~\cite{rzepkowski21}. 

In this paper we overcome the limitations of the previously discussed schemes, utilizing a higher level of control over the system, including the possibility
of tailoring the QE interaction and adjusting it within a single experiment. We study
a generic Hamiltonian  which is inspired by the effective interaction between a transmon qubit and photonic 
modes in a microwave cavity \cite{campagne20}. A similar Hamiltonian describes decoherence of trapped ions  \cite{monroe21}, or in microwave and optical cavity-QED \cite{hamsen17,haroche20,langenfeld21},
thus the type of interaction is relevant for practical implementations of quantum information
concepts.
Note, that in all of these scenarios, it is possible to control the QE interaction, 
so the proposed scheme can be realistically operated. 

The Hamiltonian under study is a slightly modified version of the spin-boson model, and QEE generated by the interaction is not detectable via the schemes of Refs \cite{roszak19a,rzepkowski21,strzalka21}. This undetectability of QEE is due
to commutation of different terms in the Hamiltonian, rather than symmetries in the density matrix (see Appendix of Ref.~\cite{rzepkowski21}). Thus the proposed QEE detection scheme is applicable to a broader class of problems, as it can be applied to any PD interaction Hamiltonian. 
The Hamiltonian used to detect the entanglement cannot commute with the initial state
of the environment, but otherwise the choice of interaction in the detection phase is flexible, and can
be tailored to the capabilities of the experimental setup. 

The paper is organized as follows. In Sec.~\ref{sec2} we rehash the formalism for PD evolutions
		and the relevant criteria for QEE. 
		The system under study is presented in Sec.~\ref{sec3}. Sec.~\ref{sec4}
		contains a description of the detection scheme and the results for transmon qubits.
		Sec.~\ref{sec5} concludes the paper.

\section{QEE for PD evolutions\label{sec2}}
Pure-dephasing evolutions, meaning joint qubit-environment evolutions which lead to pure dephasing
of the qubit after the environmental degrees of freedom have been traced out, are particularly relevant for the study of QEE, because straightforward methods for qualification of the obtained QE states
as entangled or separable exist \cite{roszak15,roszak18}. Furthermore, PD evolutions are often the dominant 
source for qubit decoherence in many solid-state scenarios. This includes bound
charge states interacting with phononic modes \cite{borri01,vagov03,vagov04,glassl13,tahara14,salamon17,seidelmann19},
as well as spin qubits interacting with spin environments under sufficiently large applied magnetic 
fields \cite{zhao12,kwiatkowski18,bartling22,bayliss22,wang22,onizhuk23}.
For the scope of this article, physical qubit realizations that allow for tuning of the interaction
Hamiltonian are of particular interest. Such can be found in hybrid systems, and particular examples
involve superconducting qubits \cite{touzard19,campagne20,gao21,ma21a,delaney22,hassani23}, the decoherence of which can effectively be described as pure dephasing,
and trapped ions \cite{leibfried03,kienzler16,lv18,bock18,landsman19,monroe21,kokail22,matsos23}.

PD evolutions are characterized by the fact that the environment correlated to the qubit cannot change the occupations of the qubit
(typically because the energy difference between the qubit states is too large in comparison with the 
energy avaliable in the environment). For an interaction Hamiltonian to be only capable of inducing such 
decoherence, it must commute with the free qubit Hamiltonian.
In such cases, the full QE Hamiltonian can be transformed into the form
\begin{equation}
	\label{ham0}
	\hat{H}=|0\rangle\langle 0|\otimes\hat{V}_0 +|1\rangle\langle 1|\otimes\hat{V}_1
\end{equation}
and the resulting QE evolution operator is given by
\begin{equation}
	\label{u}
	\hat{U}_{PD}(t)=|0\rangle\langle 0|\otimes\hat{w}_0(t) +|1\rangle\langle 1|\otimes\hat{w}_1(t).
\end{equation}
Here the operators that describe the evolution of the environment conditional on the pointer state of the qubit 
$|i\rangle$ are given by $\hat{w}_i(t)=\exp(-i/\hbar\hat{V}_i t)$.

If the initial state of the environment is a product state, the state of the qubit is a pure superposition
of the pointer states, 
\begin{equation}
\label{sigma0}
\hat{\sigma}(0)=|\psi\rangle\langle\psi|\otimes\hat{R(0)},
\end{equation}
with $|\psi\rangle =a|0\rangle+b|1\rangle$, $a,b\neq 0$,
then the if and only if condition of separability of the QE state at time $t$ is given by \cite{roszak15,roszak18},
\begin{equation}
	\label{cond}
	\hat{R}_{00}(t)= \hat{R}_{11}(t),
\end{equation}
with $\hat{R}_{ii}(t)=\hat{w}_i(t)\hat{R}(0)\hat{w}_i^{\dagger}(t)$ and $i=0,1$.
Here $\hat{R}(0)$ denotes the state of the environment which is fully arbitrary.
Note, that the QEE criterion of Eq.~(\ref{cond}) is derived exclusively from the PPT criterion
\cite{horodecki96,peres96a}
(and it is additionally proven that no bound entangled states are possible for the class of density matrices 
under study.).

An important part of the separability condition is the fact that both environmental operators $\hat{R}_{ii}(t)$
are density matrices, describing the state that the environment would have at time $t$ under the action
of Hamiltonian (\ref{ham0}), if the initial state of the qubit would be either $|0\rangle$ or $|1\rangle$.
This means that entanglement is connected with the presence of information within the environment about the state
of the qubit, while separable decoherence is not accompanied by information transfer between the two subsystems.

\section{Transmon qubit and microwave cavity \label{sec3}}
Superconducting qubits are among the most promising candidates for quantum technology
including scalable quantum computers. It
means that they are among the most developed in qubit operations, gates, and measurements executed on the qubit.
Many complex experimental protocols have been implemented on superconducting qubits. They are used in
IBM and Google quantum computers, as well as in other ventures \cite{ryan17,mi21}. 

The system under study consists of a superconducting transmon qubit interacting with a microwave cavity,
described by the effective Hamiltonian (see Supplemental Materials of Ref.~\cite{campagne20}),
\begin{equation}
\label{ham}
\hat{H}(t)=\hat{\sigma}_z\otimes\left[\left(\alpha\hat{a}^{\dagger}+\alpha^*\hat{a}\right)+\beta\hat{a}^{\dagger}\hat{a}
+\gamma\right].
\end{equation}
In analogy to the spin-boson model, the interaction is proportional to the 
$\hat{\sigma}_z$ operator on the qubit subspace, and the operators $\hat{a}^{\dagger}$, $\hat{a}$
are photonic creation and annihilation operators. The parameter $\alpha$ is the spin-boson coupling constant, $\beta$ is the dispersive shift, while the term $\gamma$ comes from the free Hamiltonian of the 
qubit. The Hamiltonian differs from the spin-boson model by the middle dispersive term, which is 
part of the interaction here, whence in the spin-boson model it would describe the free Hamiltonian of the 
environment (and be proportional to unity on the qubit side instead of $\hat{\sigma}_z$).
Note that adding a quadratic term to Hamiltonian (\ref{ham}), $\Delta\hat{a}^{\dagger 2} + \Delta^* \hat{a}^{2}$, 
would not invalidate the scheme, 
but simply require technically more involved calculations.
Although it is outside the scope of this work, the presented
scheme can be used to detect QEE for quadratic Hamiltonians of bosons coupled to the qubit.

The QE evolution operator corresponding to Hamiltonian (\ref{ham}) can formally be written in the form
given by Eq.~(\ref{u}),
with $\hat{w}_{\pm}(t)=\exp\{\mp \frac{i}{\hbar}\hat{V}t \}$, where
\begin{equation}
\label{v}
\hat{V}=\left[\left(\alpha\hat{a}^{\dagger}+\alpha^*\hat{a}\right)+\beta\hat{a}^{\dagger}\hat{a}
+\gamma\right].
\end{equation}
Thus, the conditional evolution operators of the environment, $\hat{w}_{\pm}(t)$, obviously commute
and entanglement generated via the interaction cannot be detected using any of the schemes of Refs
\cite{roszak19a,rzepkowski21,strzalka21}.

\section{Detection of QEE \label{sec4}}
\begin{figure}[!tb]
	\centering
	\includegraphics[width=0.85\columnwidth]{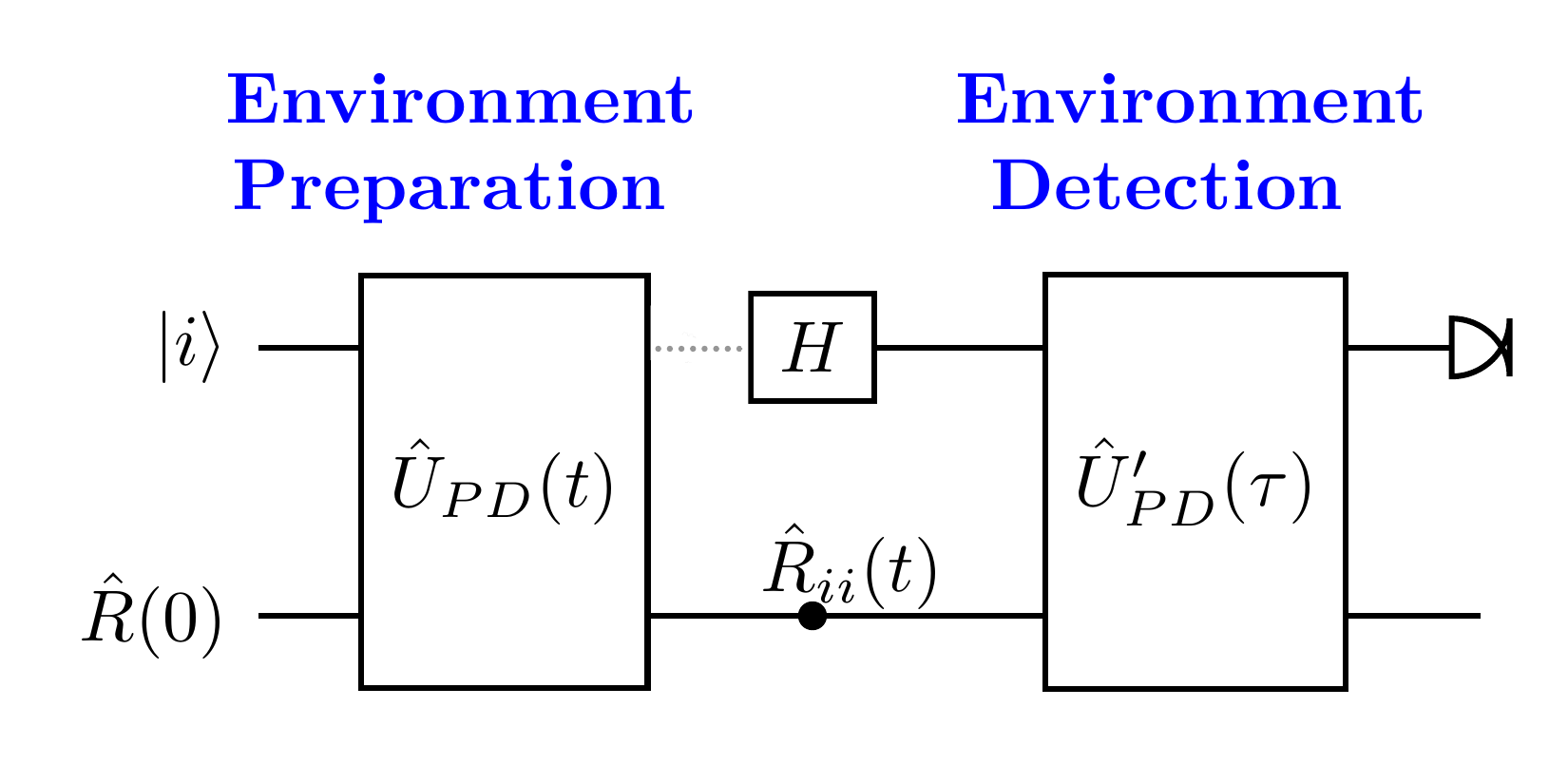}
	\caption{Circuit representing a single run of the QEE detection scheme. Initially the qubit is
		in pointer state $|i\rangle$ and environment is in state $\hat{R}(0)$.
		They interact for time $t$ preparing the environmental state $\hat{R}_{ii}(t)$, while the qubit is not affected. Then the Hadamard gate is used to prepare a superposition qubit state
		and the interaction is changed to the probe settings, followed by qubit coherence measurements
		(e.~g.~ measurements of observables defined by the Pauli operators $\hat{\sigma}_x$ and $\hat{\sigma}_y$).
		\label{g1}}
\end{figure}

The detection scheme under study is an extension of the scheme of Ref.~\cite{roszak19a};
the circuit representing the scheme is presented in Fig.~\ref{g1}.
We want to detect entanglement that is potentially generated due to the QE interaction 
governed by the Hamiltonian (\ref{ham}) after time $t$ for the qubit initially in a pure superposition of its
pointer states 
(and arbitrary initial state of the environment $\hat{R}(0)$), as in state (\ref{sigma0}).
The idea is based on the fact that when the qubit is initialized in one of the pointer states $|i\rangle$,
$i=0,1$, instead, 
the environment will evolve into the state $\hat{R}_{ii}(t)$ due to the interaction governed by Hamiltonian (\ref{ham}).
Thus environmental states $\hat{R}_{00}(t)$ and $\hat{R}_{11}(t)$ which contain information about 
entanglement (entanglement is present
at time $t$ for the initial superposition of the qubit 
if and only if $\hat{R}_{00}(t)\neq\hat{R}_{11}(t)$) can be created in a straightforward manner.
This is an indirect scheme, meaning that we are quantifying the capability of the first interaction in combination with the first environmental initial state to create entanglement. The entanglement we are studying is never present in the system-environment state. The second part of the scheme is meant to witness a difference
between the conditional density matrices in the environment, thus directly probing the if and only if condition for QEE.

The task at hand is to read out this information, but only by accessing the qubit.
To this end, the Hadamard gate is performed on the qubit at time $t$ to get an equal superposition state
(note that this will yield a superposition with a minus sign for the $|1\rangle$ state).
If $\hat{w}_0(t)$ and $\hat{w}_1(t)$ do not commute, then qubit decoherence induced by the same interaction 
Hamiltonian will  
differ depending on the new initial environmental state
$\hat{R}_{ii}(t)$ for $i=0,1$, allowing for the indirect detection of entanglement \cite{roszak19a}.
In the case under study the two conditional evolution operators commute, 
$[\hat{w}_0(t),\hat{w}_1(t)]=0$, thus detection requires a different Hamiltonian in the detection phase, such that does not commute with the initial Hamiltonian.
Note that there are multiple ways of preparing the qubit in a superposition state between the preparation and
detection stages, for example, one could measure the qubit in a basis of two states that are superpositions instead of using the Hadamard gate.

The evolution of the coherence curves after the Hadamard gate is applied is
explicitly given by 
\begin{equation}
\label{deta}
\rho_{01}^{(i)}(\tau)=\pm\frac{1}{2}\tr\left[\hat{{w}'}_0(\tau)\hat{R}_{ii}(t)\hat{{w}'}_1^{\dagger}(\tau)\right].
\end{equation}
This is the quantity measured on the qubit at the end of the procedure, as denoted in Fig.~\ref{g1}.
The ``$\pm$'' signs are the result of rotating the qubit via the Hadamard gate. 
We assume that the interaction after the Hadamard gate is still governed by a PD Hamiltonian of the same 
form as the interaction under study (\ref{ham}).
The ``prime'' in the new set of conditional operators $\hat{w}'_i(\tau)$ indicates that the parameters 
$\alpha$ or $\beta$ have been changed with respect to the original interaction Hamiltonian.
Time $\tau$ is the time elapsed after the application of the Hadamard gate.

For the scheme of Fig.~\ref{g1} to work, the Hamiltonian must be modified in such a way that each 
conditional evolution operator from the first step, $\hat{w}_i(t)$, does not commute 
with at least one operator
from the second step, $\hat{w}'_j(\tau)$.
Then we can formulate the qubit-environment {\it entanglement witness}:
if at any time $\tau$ the two coherence curves given by Eq.~(\ref{deta}) with $i=0,1$, respectively,
do not coincide,
this means that $\hat{R}_{ii}(t)\neq\hat{R}_{jj}(t)$, violating the separability criterion. 
Consequently, we know that if the qubit were prepared in any superposition
$\alpha|0\rangle+\beta|1\rangle$ instead of one of the pointer states, and evolved together with the environment initially in state $\hat{R}(0)$,
then the QE state at time $t$ would have been entangled. 

If the two curves coincide at all times $\tau$, this means that either the separability 
criterion is not violated, 
or the violation of separability has not been witnessed.
The reason is that the trace operation over the degrees of freedom of the environment
performs an average over a vast number of parameters (especially for large environments)
and certain symmetries in the QE interaction can lead to the cancellation of relevant terms, making the
result inconclusive.
In such cases, the second Hamiltonian has to be further modified.

Note, that the ``detection'' Hamiltonian, determining $\hat{U}'_{PD}(\tau)$ does not need to be of the same form as Hamiltonian (\ref{ham})
for the scheme to detect entanglement, nor does it have to be of PD type. We use a Hamiltonian analogous
to that of Eq.~(\ref{ham}) due to experimental feasibility to tuning of the interaction present for
transmon qubits and trapped ions. 

\begin{figure}[!tb]
	\centering
	\includegraphics[width=0.66\columnwidth]{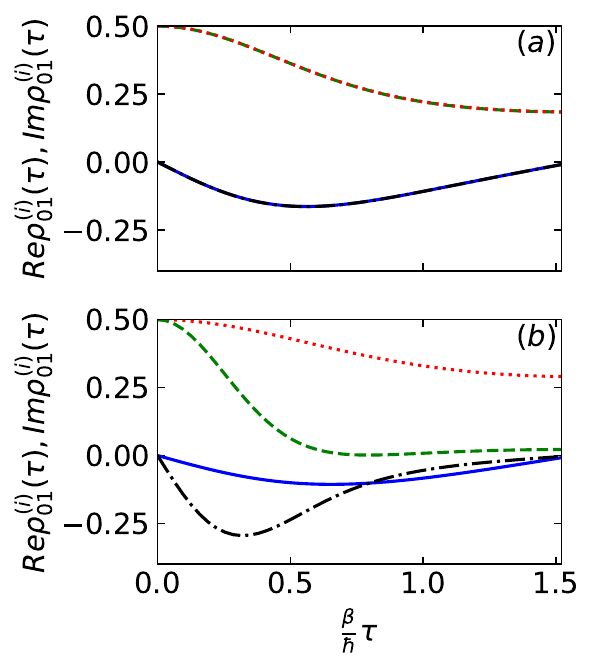}
	\caption{Evolution of the real and imaginary parts of qubit coherence
		[given by Eq.~(\ref{deta}) and corrected for the minus sign stemming from the Hadamard gate]
		for the measurement
		part of the QEE detection scheme, Fig.~\ref{g1}, at zero temperature. The preparation time is
		(a) $\beta t/\hbar=0$ (separable state at time $t$) and (b) $\beta t/\hbar=2$
		(entangled state at time $t$).  
		Real part of coherence (upper curves): $i=0$ - dotted, red lines; $i=1$ -
		dashed green lines. Imaginary part of coherence (lower curves):
		$i=0$ - solid blue lines;  $i=1$ - dotted-dashed black lines.
	The Hamiltonian parameters are set to $\alpha/\beta=(1+i)/2$ in the preparation part, and $\alpha/\beta=1/\sqrt{2}$ in the measurement part.
Any discrepancy between the two upper curves or the two lower curves signifies entanglement, thus entanglement is witnessed in panel (b). \label{fig_d2}}
\end{figure}

\begin{figure}
	\centering
	\includegraphics[width=1.\columnwidth]{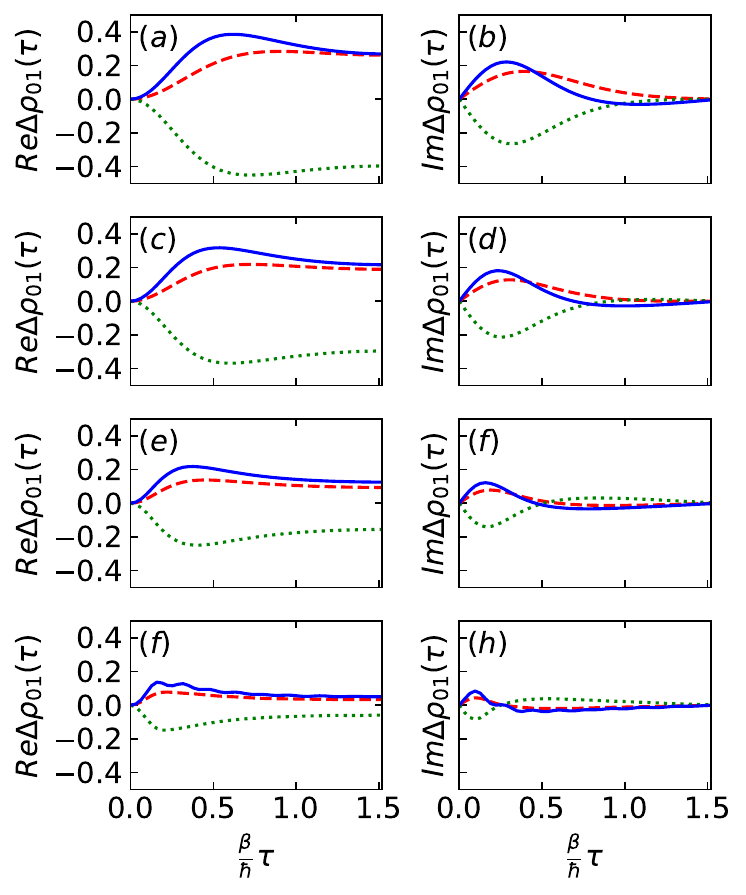}
	\caption{Evolution of the difference of qubit coherence 
		in the measurement stage of the QEE detection scheme; real part is on the left and 
		imaginary part on the right. Different curves correspond
		to different preparation times: dashed red lines for $\beta t/\hbar =\pi/6$, solid blue lines for $\beta t/\hbar=2$, and dotted green lines $\beta t/\hbar=3\pi/2$.
		Different panels correspond to different effective  temperatures: (a),(b) $k_BT/\Gamma =0$, (c),(d) $k_BT/\Gamma =0.5$, (e),(f) $k_BT/\Gamma =1$, (g),(h) $k_BT/\Gamma =2$. The Hamiltonian parameters are set to $\alpha/\beta=(1+i)/2$ in the preparation part, and $\alpha/\beta=1/\sqrt{2}$ in the measurement part. }\label{fig_d1}
\end{figure}

In the following, we aim to indirectly detect the formation of QEE for Hamiltonian (\ref{ham})
with $\alpha/\beta=(1+i)/2$ and initial state of the environment 
given by the thermal equilibrium state \cite{glauber63}
with the density matrix 
$\hat{R}(0)\sim\exp\{\Gamma\hat{a}^{\dagger}\hat{a}/{\mathrm{k_B}T}\}$.
From Ref.~\cite{strzalka24}, we know that the evolution is entangling at finite temperatures.
The measurement phase involves the same Hamiltonian with a different
parameter $\alpha$,
$\alpha/\beta=1/\sqrt{2}$.
The parameters $\alpha$ and $\beta$
in both phases are chosen in a fairly arbitrary way, such that entanglement can be detected and
the effect of the interaction is visible on similar time-scales in both phases. We have found that an imaginary 
component in only one of the Hamiltonians tends to maximize the discrepancy between the two types of decoherence 
curves (\ref{deta}).

Fig.~\ref{fig_d2} shows the zero-temperature
evolution of the real and imaginary parts of the transmon qubit 
coherence in the measurement part of the time-dependent QEE detection scheme, for the qubit prepared in pointer state $|0\rangle$ and 
prepared in pointer state
$|1\rangle$, given by Eq.~(\ref{deta}). 
In Fig.~\ref{fig_d2} (a) the preparation time is zero, so there is no QEE to be detected; the detection curves 
here overlap exactly.
In Fig.~\ref{fig_d2} (b)
the preparation phase lasted $\beta t/\hbar=2$, and any difference in either the real or the imaginary parts signifies that the Hamiltonian with $\alpha/\beta=(1+i)/2$ would lead to QEE between
any superposition of qubit pointer states and the zero-temperature Gibbs state
after this time has elapsed. We observe a large
difference between the coherence curves both in the real and imaginary parts, so this method not
only allows for the detection of QEE, but this entanglement would be relatively easy to distinguish
in the data.

Fig.~\ref{fig_d1} contains the differences between the $\tau$-dependent coherence curves 
introduced in Fig.~\ref{fig_d2}, given by
\begin{equation}
	\label{signal}
\Delta\rho_{01}(\tau)=\rho_{01}^{(0)}(\tau)+\rho_{01}^{(1)}(\tau).
\end{equation}
The formula takes into account the sign difference introduced into the states of Eq.~(\ref{deta})
via the Hadamard gate, so Eq.~(\ref{signal}) is the difference in qubit coherence for the situation, when 
both superpositions in the detection phase were initialized in the $|+\rangle=\frac{1}{\sqrt{2}} \left(|0\rangle+|1\rangle\right)$
state.
The left column contains the difference between the real parts (depicted in the left column)
and the imaginary parts are (right column). Each plot contains three curves
corresponding
to different preparation times $t$: $\beta t/\hbar =\pi/6$, $\beta t/\hbar=2$, and $\beta t/\hbar=3\pi/2$,
so they detect entanglement for different moments of the evolution driven by the original 
Hamiltonian. 
Temperature is increased from top to bottom.
Depending on the time at which entanglement is probed, the difference between the decoherence 
curves can be very significant (which is to be expected, as there exist moments in time during
the evolution when there is no QEE and the curves would be the same).
The signal, Eq.~(\ref{signal}), decreases with increasing temperature which is related to the fact that the amount
of generated entanglement is inversely proportional to temperature.

Note that the obvious choice of Hamiltonian for the measurement phase, namely Eq.~(\ref{ham}) with $\alpha=0$,
does not yield any difference between qubit coherence evolutions and thus does not signify the presence
of entanglement in the system. This is the case even though the different conditional evolution operators
of the environment do not commute, and is a consequence of the symmetries present in the initial state
of the environment. 

\section{Conclusion\label{sec5}}
We generalized
the scheme for indirect QEE detection for it to detect entanglement in PD evolutions \cite{roszak19a}
that is in principle not 
detectable by operations and measurements on the qubit alone. To this end we newly took advantage of
the possibility to tune the interaction between qubits and their environment.
A leading example of such systems (with entanglement undetectable by standard schemes and tunable interactions) is a transmon qubit interacting with a microwave cavity.
We present results for the operation of the entanglement detection scheme for such qubits,
demonstrating an efficient witness that should be detectable even at finite temperatures.

This removes the fundamental 
limitation for QEE detection, which is related to essential commutation of different parts of the interaction Hamiltonian 
and to symmetries present in the initial state of the environment. Both obstacles can be overcome as long
as there is control over the interaction. The choice of concrete Hamiltonian which should be used for the 
measurement is flexible, and should be determined on an individual basis, depending on the properties
of the platform under study including the limitations on the control over the interactions.

\section*{Acknowledgment}
KR and MS: This project is funded within the QuantERA II Programme that has received funding from the EU H2020 research and innovation programme under GA No 101017733, and with funding organization MEYS Czech Republic.
MS acknowledges support by the Grant Agency of Charles University (122124).
RF has been supported by the project GA22-27431S of the Czech Science Foundation, by the project No. LUC25006 of MEYS Czech Republic, and by project CZ.02.01.01/00/22$\_$008/0004649 (QUEENTEC) of EU and MEYS Czech Republic. Computational resources were provided by the e-INFRA CZ project (ID:90254),
supported by the Ministry of Education, Youth and Sports of the Czech Republic.

%

\end{document}